\documentclass[preprint,amssymb,showpacs,supersriptaddress,floatfix]{revtex4-2}
\usepackage{graphicx}
\usepackage{dcolumn}
\usepackage{bm}
\usepackage[utf8]{inputenc}
\usepackage[T1]{fontenc}
\usepackage{mathptmx}
\usepackage{etoolbox}
\usepackage{float}
\usepackage{amsmath,amssymb}
\usepackage[dvipsnames]{xcolor}
\usepackage{subcaption}
\makeatletter
\def\@email#1#2{
 \endgroup
 \patchcmd{\titleblock@produce}
  {\frontmatter@RRAPformat}
  {\frontmatter@RRAPformat{\produce@RRAP{*#1\href{mailto:#2}{#2}}}\frontmatter@RRAPformat}
  {}{}
}
\makeatother

\usepackage{color}
\newcommand{\change}[1]{{\color{red}{#1}}}
\newcommand{\changeA}[1]{{\color{blue}{#1}}}
\newcommand{\changeD}[1]{{\color{green}{#1}}}
\usepackage[utf8]{inputenc}
\usepackage[T1]{fontenc}

\begin{document}



\title{Oxygen vacancies kinetics in TaO$_{2-h}$/Ta$_2$O$_{5-x}$   memristive interfaces}
\author{C. Ferreyra$^{1}$}
\author{R. Leal Martir$^{2,3}$}
\author{D. Rubi $^{1}$}
\author{María José Sánchez$^{2,3}$}

\affiliation{$^{1}$Instituto de Nanociencia y Nanotecnología (INN), CONICET-CNEA, nodo Buenos Aires, Argentina.\\
$^{2}$Instituto de Nanociencia y Nanotecnología (INN),CONICET-CNEA, nodo Bariloche,8400 San Carlos de Bariloche, Río Negro, Argentina.\\
$^{3}$Centro Atómico Bariloche and Instituto Balseiro (Universidad Nacional de Cuyo), 8400 San Carlos de Bariloche, Río Negro, Argentina.}

\email{majo@cab.cnea.gov.ar}

\begin{abstract}



Oxygen vacancies (OV) are pervasive in metal oxides and play a pivotal role in the switching behaviour of  oxide-based memristive devices.
In this work we address, through a combination of experiments and theoretical simulations, OV dynamics in Pt/TaO$_{2-h}$/Ta$_2$O$_{5-x}$/TaO$_{2-y}$/Pt devices. In particular, we focus on the  RESET transition (from low to high resistance), induced by  the application of electrical pulse(s), by choosing different initial OV profiles and studying their kinetics during the mentioned process. We demonstrate that by selecting specific OV profiles it is possible to tune the characteristic time-scale of the RESET. Finally, we show that the implementation of gradual RESETs, induced by  applying many (small) successive pulses, allows estimating the activation energies involved in the OV electromigration process. Our results help paving the way for OV engineering aiming at optimizing key memristive figures such as switching speed or power consumption, which are highly relevant for neuromorphic or in-memory computing implementations.

\end{abstract}

\maketitle


Recent advances in big data manipulation and  machine learning demand for low-power in-memory computing operations. Despite substantial progress in software and processing algorithms centered on artificial neural networks \cite{lecun_2015,   zhang_2020, ning_2023}, the evolution of hardware architectures aiming to avoid the Von Neumann bottleneck 
continues to encounter substantial challenges. In this scenario,
 electronic hardware that could mimic the structure and information processing  mechanisms of biological systems is emerging as a promising alternative \cite{upadhyay_2019, kendall_2020,zhang_2020}. 
In particular memristors are playing a essential role \cite{val_2013, kumar_2022}, contributing to the realization of artificial synapses and in-memory processing and aiming to  eliminate the physical separation between memory and processing units characteristic of the usual Von Neumann architectures \cite{li_2018,mehonic_2020, sebastian_2020, yao_2020, sun_2021, zhou_2022}.
 Memristors are metal/insulator/metal capacitor-like structures that exhibit the resistive-switching  (RS) phenomena, i.e. the reversible change of the electrical resistance upon the application of a electrical stress \cite{saw_2008,yan_2008,waser_2010,borghetti_2010, pan_2014,wang_2020}.

In Non-Volatile RS (NVRS) the resistive states persist in time after the electrical stimulus is turned off. Hysteresis current–voltage switching loops are a fingerprint of this behavior. On the other hand, in Volatile RS (VRS) the  attained  resistive states  relax  once the electrical stimulus is turned off \cite{zhuo_2013,zhou_2022}. Thus, several synaptic and neuron functionalities can  be electrically emulated based on the lifespan of the resistance states. In the past years, extensive demonstrations of  RS  in oxides  were realized, mostly on binary oxides such TiO$_2$,  HfO$_2$, and Ta$_2$O$_5$ \cite{lee_2011, gale_2014,kumar_2022,xiao_2023}. 
In particular Ta$_y$O$_x$ based devices stand out among others, exhibiting exceptional figures as  ON-OFF ratios up to $10^5$,  retention times above 10 years and endurances of up to $10^{10}$ cycles \cite{lee_2011,breuer_2016}. RS has been reported as bipolar and its origin related to the formation and retraction/rupture  of nanoscale conducting oxygen vacancies (OV) filaments upon  polarity reversal of the applied stimulus, ruling the SET (high to low resistance) and RESET  (low to high resistance) transitions respectively \cite{bae_2012, chen_2015, park_2015,huang_2018, kim_2020}.


The development of neuromorphic devices based on memristors  requires a  comprehensive knowledge  of the mechanisms ruling the RS. Understanding the role of OV and  their dynamics  is crucial for extensive optimization (improve the switching speed, minimize the energy consumption, enlarge the endurance) and  design of  oxide-based memories for neuromorphic applications. Numerous reports show that OV,  present as extrinsic or intrinsic defects, play a  central role in  the RS mechanism \cite{waser_2010,bae_2012,balatti_2013}. 
In particular the RESET process in TaO$_x$ cross-bar arrays was recently analyzed employing time-resolved pulsed experiments and attributed to the OV motion  governed by drift and diffusion processes \cite{marche_2016}.
Despite the obtention of qualitative knowledge, it remains unclear in which way specific OV profiles and their kinetics affect the RS properties of the devices under electrical bias.
In addition, the intricate dynamics involved in OV transport  remains incompletely elucidated. Thus, obtaining evidence pertaining to the motion of OV  proves challenging and usually involve complex in-operando experiments \cite{herpers,nukala_2021,jan_2023}.
In order to pave the way along this central issue, in this work we investigate- using a much simpler experimental approach- the transient electrical response in a Pt/TaO$_{2-h}$/Ta$_2$O$_{5-x}$/TaO$_{2-y}$/Pt device during the RESET process.

Assisted by the Voltage-Enhanced Oxygen Vacancy model (VEOV) \cite{roz_2010,ghe_2013, Fer_2020, LMartir_2023}, we identify the specific OV profile at the origin of the RESET transition and we determine its time evolution, providing an unique insight into OV dynamics along this process.  In addition, through the study of successive partial RESETs, we were able to estimate the activation energy for OV migration, which is a key issue for controlling the RS effect and designing efficient operation electrical protocols.


\subsection{EXPERIMENTAL SET UP}

\begin{figure}[h!]
    \centering
    \includegraphics[width=0.8\linewidth]{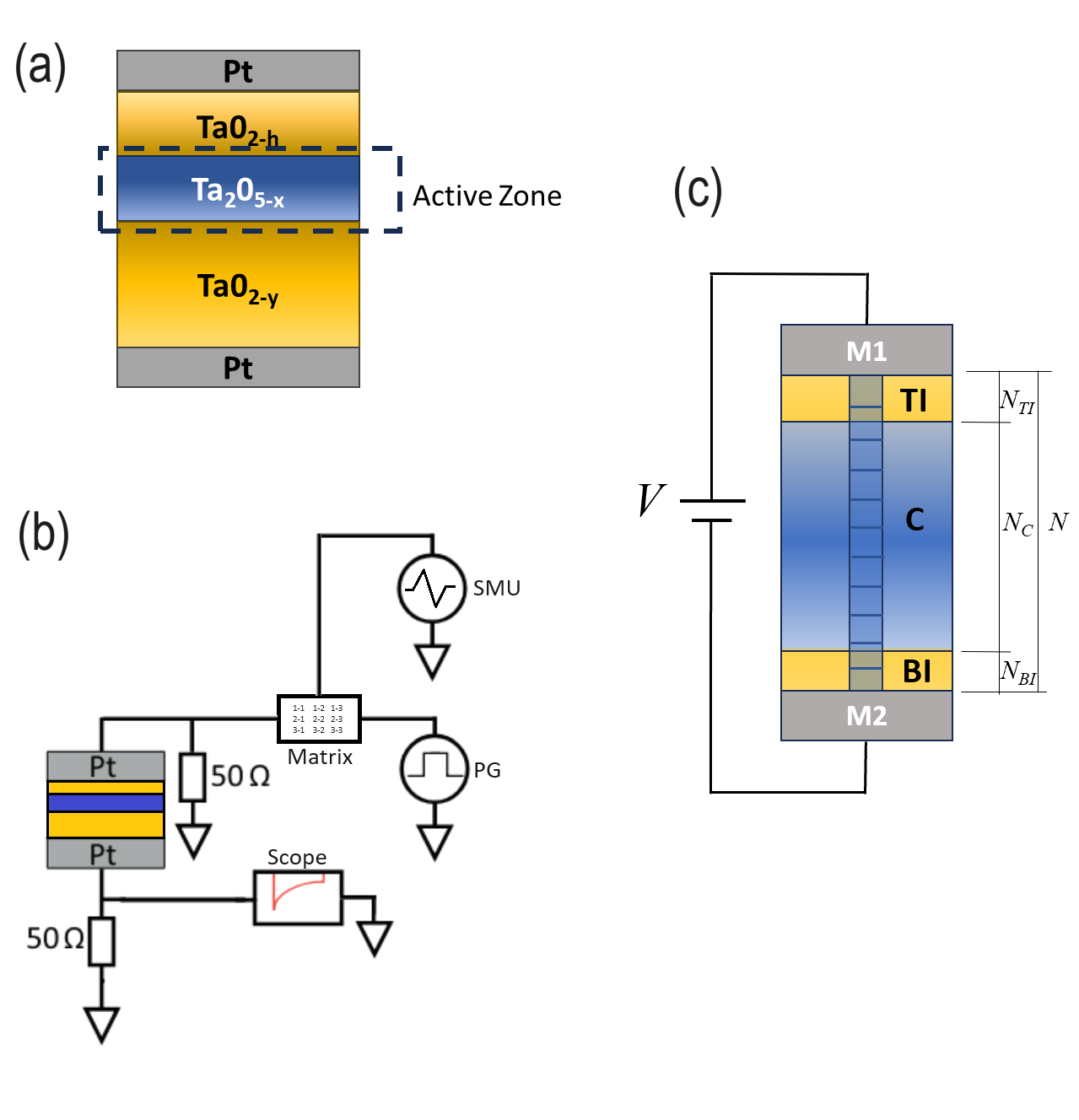}
    \caption{ (a) Sketch of the post formed device. The region enclosed by the  dashed rectangle comprises the  Ta$_2$O$_{5-x}$ layer and the two active interfaces, TaO$_{2-h}$/Ta$_2$O$_{5-x}$ (TI) and Ta$_2$O$_{5-x}$/TaO$_{2-y}$ (BI) respectively. (b) Experimental set up employed for the electrical characterization. (c) Sketch of the device and the active zone employed in the numerical modelling. See text for more details}
    \label{fig:f1}
\end{figure}

A bilayer Ta-oxide thin film was deposited on commercially platinized silicon using Pulsed Laser Deposition (PLD) -the laser fluence was fixed at 1.5 J cm$^{-2}$- from a Ta$_2$O$_5$ ceramic target at room temperature. The deposition was made in 2-steps,using oxygen pressures of 0.01 and 0.1 mbar, respectively. We obtained two layers, a more reduced with a thickness of 35 nm and a more oxidized with a 15 nm thickness, as shown in the sketch of Fig. \ref{fig:f1}(a). It is usually considered that more reduced layers behave as OV source/sink that favors the reduction/oxidation
of more oxidized ones, driving the memristive effect \cite{lee_2018}.

Top Pt electrode was made by sputtering  and micro-structured using optical lithography  \cite{Fer_2020}. On the other hand, the Pt of the substrate acts as the bottom electrode, which was grounded. The electrical
stimulus (voltage) was applied to the top Pt electrode.



The pristine device starts in  a low resistance state of  $\approx$ 100 $\Omega$. By applying a negative stimulus  of - 3V during 1 ms, an intermediate resistance state of the order of k$\Omega$ was attained, awakening the memristive response.
The electroformed device results in the stack
   Pt/TaO$_{2-h}$/Ta$_2$O$_{5-x}$/TaO$_{2-y}$/Pt \cite{Fer_2020}, as sketched  in Fig.\ref{fig:f1} (a). The active   zone  for RS comprises the  more oxidized layer, Ta$_2$O$_{5-x}$, sandwiched between  the two reduced layers, defining a quasi-symmetric configuration with two  active memristive interfaces: the top interface (TI) TaO$_{2-h}$/Ta$_2$O$_{5-x}$ and the bottom interface (BI) Ta$_2$O$_{5-x}$/TaO$_{2-y}$, respectively. On the other hand  Pt/TaO$_{2-h(y)}$  is an ohmic interface playing a passive role in the RS. 
   The memristive behavior  of the device was studied in detailed in Ref.\onlinecite{Fer_2020} and can be understood in terms of OV electromigration between  the central zone, Ta$_2$O$_{5-x}$, and  the interfacial regions under the applied stimulus.

The electrical characterization was performed in two steps. Firstly, using a Keithley 2612 source-meter, we applied the RS protocol (described below in Sec.\ref{sr}) to check the correct RS behavior and tune the desired resistive state of the sample. Then, with an Agilent 81110A pulse generator, a single RESET pulse is applied. A homemade matrix allowed to select the needed instrument in each case. The RESET current response was measured using a shunt connected to a Tektronix TDS5032B Oscilloscope. The  experimental setup is shown in Fig.\ref{fig:f1} (b).


\subsection{RESULTS}
\label{sr}

\begin{figure}[h!]
    \centering
    \includegraphics[width=1\linewidth]{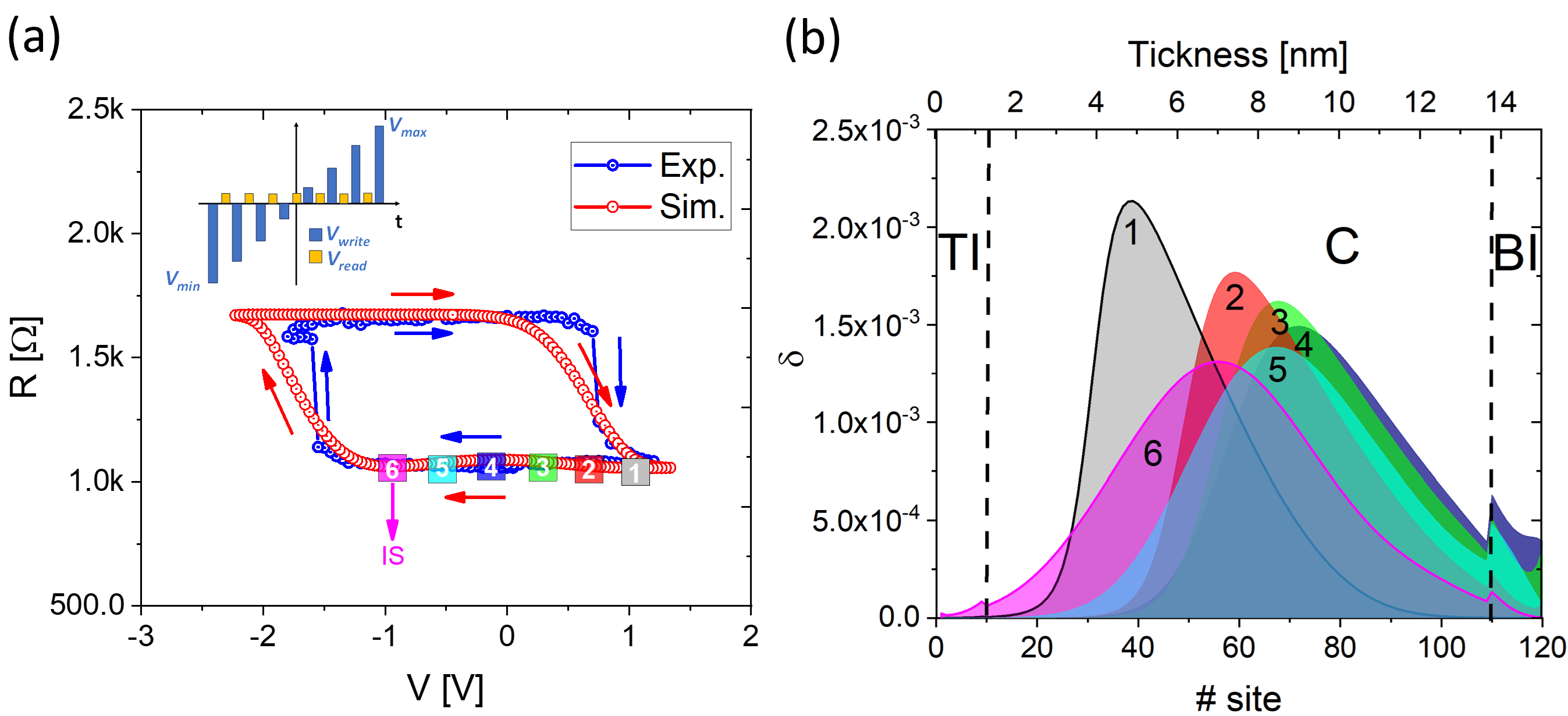}
    \caption{(a) Clockwise Hysteresis switching loop (HSL): Blue dots (experimental), red dots (simulations). The arrows indicate the circulation of the HSL. The  coloured squares numbered $1-6$   refer to selected values of  resistances used for the analysis of the RESET transition. Inset: experimental electrical protocol displaying  the write and read pulses respectively.
    (b)   OV distributions obtained from the VEOV model along the active region comprising the TI, C and BI, for each of the  resistance values labelled $1-6$  in the HSL of panel (a). The vertical dashed lines indicate the boundaries of the TI and the BI respectively. See text for more details.}
    \label{fig:f2}
\end{figure}

The RS protocol consists of a pulsed ramp that alternates between write pulses, which changes the resistance state of the device, and read pulses that capture their remanent resistance states (see inset of Fig.\ref{fig:f2} (a)).
From these pulses, we obtain respectively, the dynamic I-V curves (not shown) and the so-called Hysteresis Switching Loop (HSL) \cite{roz_2010} which displays the remanent resistance as a function of the applied (write) voltage. Depending on the polarity and the symmetry of the voltage protocol, defined by  the maximum and minimun intensities  $V_{max}$ ($\ge$0) and $V_{min}$ ($\le$0),  different HSLs with  shapes and chiralities -reflecting the activation of one memristive interface, the other or both- can be obtained \cite{Fer_2020}.
For the present device, the HSLs presents a squared-like shape with a clockwise (CW) chirality for $|V_{max}| <|V_{min}|$, and  a counter-clockwise (CCW) chirality for  the inverted  voltage excursion i.e  
$|V_{max}| >|V_{min}|$.
In addition, when the voltage excursions are enlarged and symmetrized to $V_{max} =-V_{min}$, the  HSL displays a table with legs  (TWL) -like shape \cite{Fer_2020}.
The squared HSLs  correspond to a single active interface while the TWL   corresponds to two (complementary) active interfaces; that is, when one switches from low to high resistance the other one does it  inversely \cite{roz_2010}. 
Therefore,  each memristive interface can be selectively activated or deactivated  according to the polarity and intensity of the applied stimuli \cite{Fer_2020,LMartir_2023}.


Starting with a post-forming device, we prepared the device in a low resistance state by following the CW HSL, using a non-symmetrical protocol with $V_{min}=-1.8 $V and $V_{max}= 1.2$ V. The sample  was initially cycled several times to ensure stability and repeatability of the CW  HSL, displayed in  Fig.\ref{fig:f2} (a) (blue dots), with  well-defined LRS ($R_L=$1.05 k$\Omega$), HRS ($R_H=$1.65 k$\Omega$) and SET and RESET  voltages, respectively. Notice that the LRS remains almost flat for  voltages in the range [1.2 V, -1.2 V],
reflecting the time-stability (non-volatility) of the resistance states.
We start  by  analysing  the  RESET process under the  application of  single electrical pulses of different intensities. 
To this end, the device was prepared in the low-resistance state (LRS) just before reaching the RESET threshold.
In particular, we  chose as the initial state (IS) the resistance $R_{IS}=$ 1.05 k $\Omega$, labelled $\#6$ (magenta square) in the CW HSL of Fig.\ref{fig:f2} (a).

\begin{figure}[h!]
    \centering
    \includegraphics[width=1.1\linewidth]{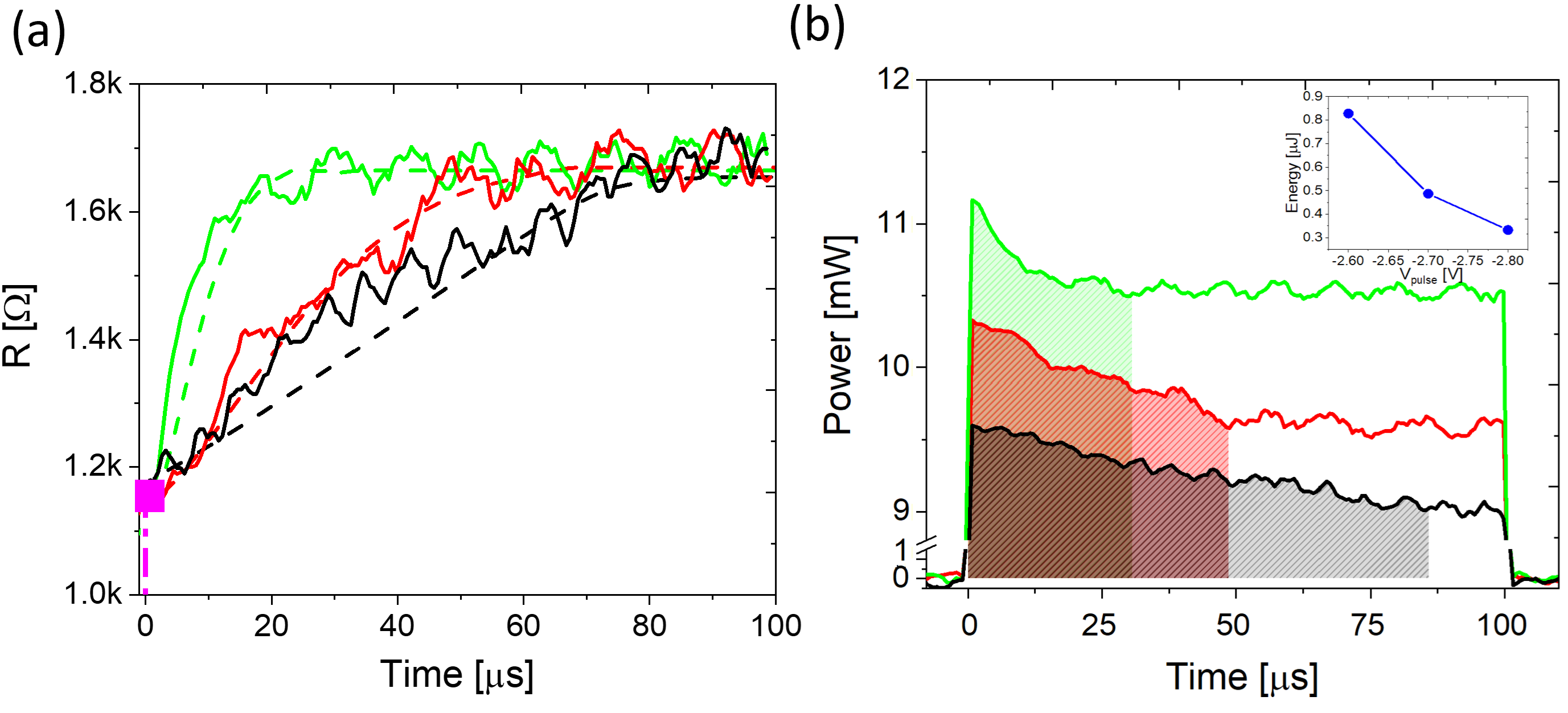}
    \caption{(a) Resistance vs. time   for different single negative bias pulses of 100 $\mu$s, starting from the IS (magenta square). Black line: -2.6 V, red line: -2.7 V, green line: -2.8 V. Experimental data are shown with solid lines, and simulations  with dashed lines.
    The complete RESET is  finished when the HRS, $R_H\approx$1.65 k$\Omega$, is attained. (b) Power consumption as a function of time. Different curves correspond to voltages amplitudes  -2.6V (black line ), -2.7V (red line ) and -2.8V (green lines), respectively. Inset: Associated energy consumption (shaded area below each of the previous curves). See text for details.}
    \label{fig_reset}
\end{figure}

Figure \ref{fig_reset}(a) shows, in continuous lines, the evolution of the resistance with time 
during the RESET process for three  different voltages applied during  100 $\mu$s, starting from the chosen IS. 
 A higher voltage induces a complete RESET attaining the HRS  in a shorter time, as expected. For -2.8 V,  it takes $\approx$ 30$\mu$s to complete the RESET process, while for -2.6 V it takes $\approx$ 100$\mu$s. 
This response is   consistent with the fact that for negative voltages, OV -as positive defects-  move from the central region towards the TI (TaO$_{2-h}$/Ta$2$O$_{5-x}$), which is active throughout the entire (RESET) process.
Thus, for higher  applied voltages,   OV  (electro) migration is enhanced  and therefore  the RESET takes place  in shorter time scales.
 In addition,  the non-linearity of the resistance as a function of time during the RESET increases with  the intensity of the  voltage - a characteristic of activated processes previously reported \cite{schroeder_2010, menzel_2015}-.

 It is interesting to estimate and compare the energy consumption involved in the RESET processes. To this end, we followed  Refs. \onlinecite{fer_2019,Fer_2020} and compute  in Fig.\ref{fig_reset}(b) the power consumption as $P=V^2/R$, where $R$ values correspond to the measured  resistances displayed in Fig.\ref{fig_reset}(a).
   In addition,  the inset shows the energy (computed as the shaded area under each of  the curves when the power variation is below 5$\%$) for the three voltage values. 
It is noteworthy that the energy involved in the RESET process is lower for higher  pulses intensities, even though  the power consumption is larger.  This is a clear evidence of the non linear dependence of the RESET process  with the applied voltage, in agreement with  the voltage-time dilemma typically present in RS devices \cite{menzel_2015} . 

To model the experimental response we  used the  1d Voltage-enhanced Oxygen Vacancies Oxygen (VEOV) migration model \cite{roz_2010, ghe_2013, Fer_2020, LMartir_2023} that describes the resistive change in oxide based memristors  as a result of the migration of the ubiquitous OV under the application of external electrical stimuli. 

The VEOV model considers a 1d chain of N nanodomains which  
 can contain  a  specific total  amount of OV. Each domain defines a site in the chain and  has  an associated resistivity, which depends locally on the OV content.  In binary oxides of transition metals, OV act as n-type dopants supplying free electrons to the conduction band and thereby  reducing  the electrical resistance. In agreement with this, we define the  resistivity of each nanodomain $i$ as   
$\rho_i=\frac{\rho_0}{(1+ A_{\alpha} \delta_i)}$, being $\delta_i$ the OV concentration at site $i$ (i.e. the ratio between the number of OV at site i over the total amount of OV),  $A_{\alpha}$ is  a parameter that depends on the zone where the domain is located (i.e. interface or not-interface zones) and $\rho_0$ is the resistivity of the oxide in the absence of OV \cite{ghe_2013}.
The total resistance of the device is computed as $R=c\sum_{i=1}^{N}\rho_i$ with c being a geometric pre-factor.



For the present case, and due to the aforementioned   interfaces,  the chain of nanodomains is divided into three regions: TI  and BI  (associated with the interfaces) which  allocate $N_{TI}$  and $N_{BI}$ sites respectively  and   region C  that represents the bulk zone of the device, as  depicted in  Fig. \ref{fig:f1} (c). 
For a given external  stimulus $V$, at each simulation step  the OV concentrations  are updated with the transfer rate $p_{i,i\pm 1}=\delta_i (1-\delta_{i\pm 1})\exp({(-V_0 +\Delta V_i)}/{(KT)})$, from site $i$ to site $i\pm 1$,
which is proportional to the OV concentration $\delta_i$ at site $i$ and to the availability of the  neighbouring sites $i \pm 1$. 
The exponential Arrhenius factor contains- given in units of thermal energy $KT$- the activation energy for OV diffusion $V_0$ (that could vary along different regions of the chain) and the local potential drop at each site, $\Delta V_i \propto V \rho_i/R$, due to the external electrical stimulus. With the  new  OV's concentrations, the local resistivities and   the total resistance are therefore updated.
In order to emulate the electrical response of the device during the RESET we established an equivalence between  experimental and numerical parameters (see the Suppl. Mat.). 

Following the described procedure we computed the remanent resistance as a function of $V(t)$ and obtained the  HSL depicted (in red) in Fig.\ref{fig:f2} (a), which  is in  good  agreement with the experimental (in blue) one.
In particular,  the simulated HRS and LRS are  almost indistinguishable from the experimental values.

A key outcome  of the VEOV model are the OV profiles associated  to  the different resistance states \cite{roz_2010,ghe_2013,ghenzi_2014,Fer_2020,LMartir_2023}. 
In particular, each of the  states labelled from  1 to 6  in Fig.\ref{fig:f2}(a) has associated the OV profiles displayed in Fig.\ref{fig:f2}(b)-labelled with the same numbers as their resistance counterparts.
These profiles could  be  taken as different  initial condition to simulate the RESET process by applying a single electrical pulse.
However, as we already mentioned, the  experiment was performed starting from the IS which, according to the VEOV model, has associated the OV profile  6 in Fig. \ref{fig:f2} (b). The simulated RESET process for this IS is  displayed in Fig.\ref{fig_reset} (a) in  dashed lines    for the three voltages considered. The agreement with the experimental results is quite good and, in particular, the simulations follow  nicely the experimental curves for $V$= -2.7 and -2.8 V. We notice that  IS profile 6 is Gaussian-like,  providing a clue about the  actual OV distribution prior to the RESET transition.

\begin{figure}[h!]
\centering
\includegraphics[width=1.1 \linewidth]{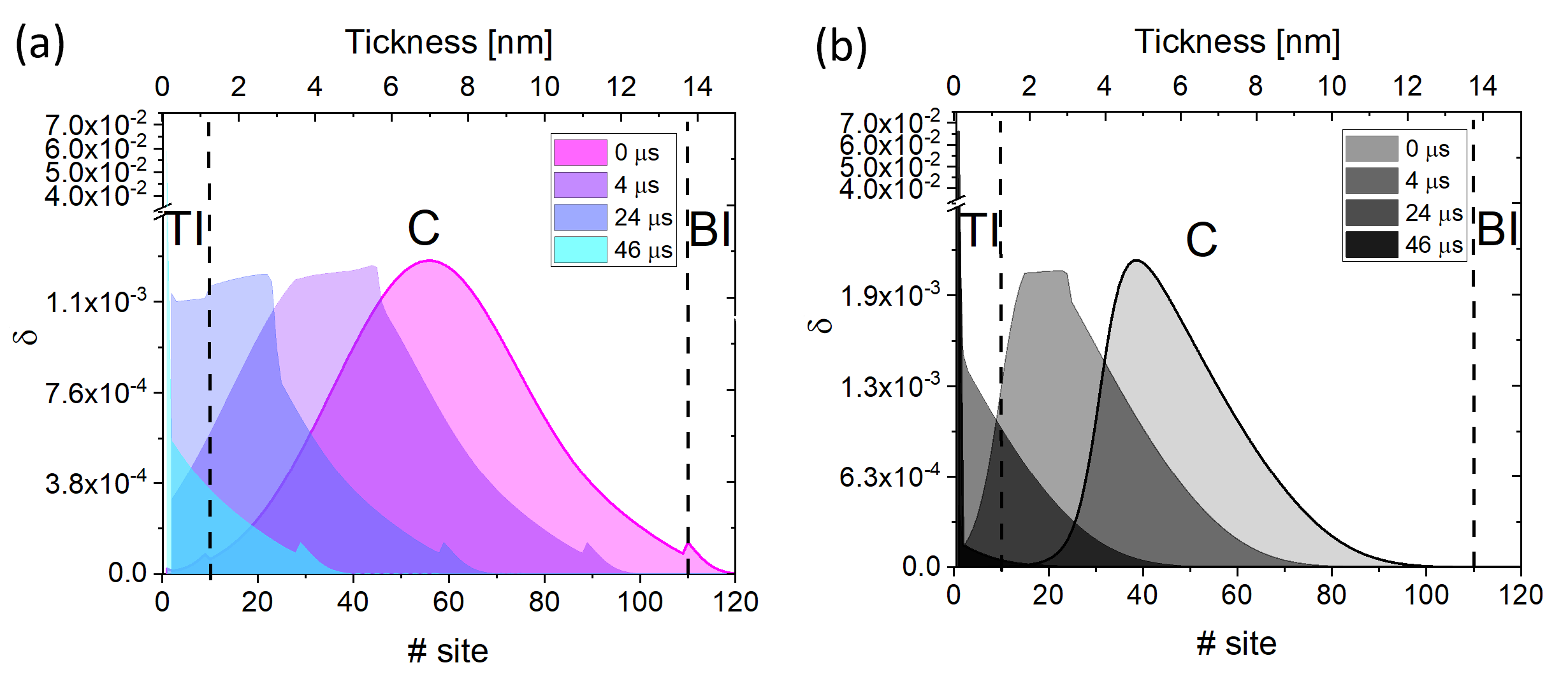}
\caption{ Snapshots of the OV´s  profiles  along the simulated RESET process for $V$=-2.7 V and for times 0,  4, 24, and  46 $\mu s$.(a) Starting from profile 6 (IS). (b) Starting from  profile 1. }
    \label{fig_ovprof}
\end{figure}

\begin{figure}[h!]
    \centering    \includegraphics[width=0.8\linewidth]{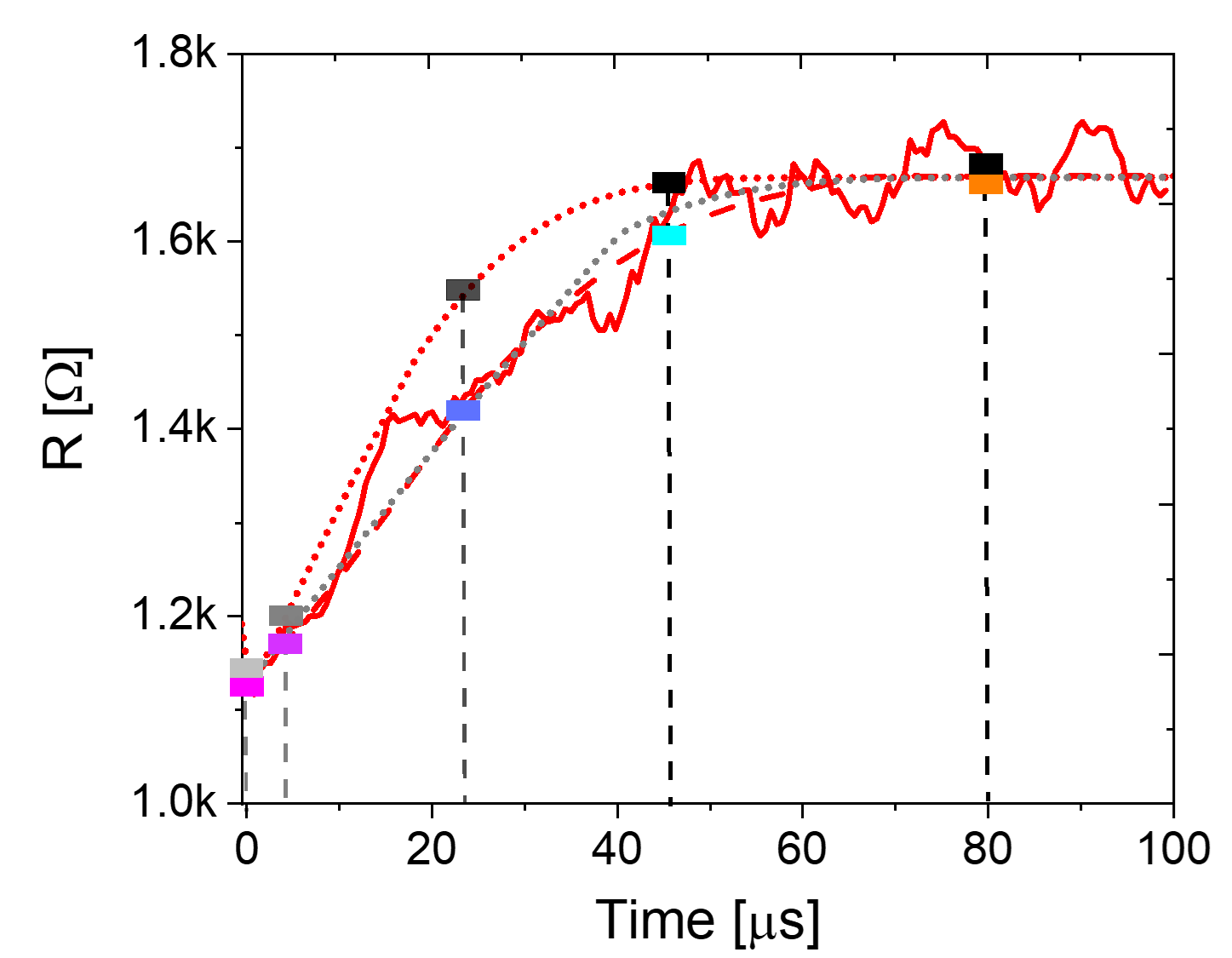}
    \caption{Experimental (red solid curve) and simulated RESET evolution  starting from the IS profile 6 (red dashed line) and from profile 1 (red dotted line), in all the cases for $V=$ -2.7V.  The grey-dotted line shows the simulated  RESET evolution starting from profile 1 but for $V$=-2.65 V.
    Square symbols indicate different resistance states, whose   associated  OV profiles-excepted  for time 80 $\mu s$- are depicted  respectively  in Fig.\ref{fig_ovprof} (a) and (b).}
    \label{fig_resim}
\end{figure}

Figure \ref{fig_ovprof} (a) shows the OV profiles for the IS (at time 0 $\mu s$)  and  for times  4, 24, and  46 $\mu s$, selected along the simulated RESET displayed in Fig.\ref{fig_resim} (red-dashed line), for $V=$ -2.7 V.
Notice that, as time evolves, OV profiles move to the TI  and accumulate there, acquiring  a shape that strongly departs from the initial Gaussian-like shape, as expected for a non diffusive dynamics driven by an external field. In particular, the OV profile for 80 $\mu s$ (not shown) -once the complete RESET is attained- is extremely sharp and localized along the first few sites. 

For the sake of comparison,  we plot in  Fig. \ref{fig_ovprof} (b) the OV profiles obtained from the simulated RESET  for $V$ =-2.7 V - displayed in  Fig.\ref{fig_resim} (red-dotted lines)-  starting from the low resistance state  labelled 1 in Fig.\ref{fig:f2} (a). 
Despite the fact that the initial and final resistance values for profiles 1  and 6 are  barely distinguishable , their transient dynamics are clearly  different.  Profile 1 completes the RESET faster than   profile 6, as   depicted   in   Fig. \ref{fig_resim}. For profile 1, the  simulated RESET  is completed in 46 $\mu s$, while neither the experimental  data (red solid red) nor the simulated evolution starting from the IS (red dashed line) -which nicely follows the experimental curve- have   attained the RESET for that time. This is  a direct consequence of the fact that the initial  OV distribution for profile  1 -which is  slightly non-Gaussian- has its maximum  closer to the TI than the IS profile 6. 
Therefore,  for profile 1 it should be possible, by  lowering the external stimulus, to slow OV dynamics in order to  have a resistance vs. time evolution that matches the one obtained for the initial IS stimulated with a  $V=$-2.7V pulse.
Indeed,  we depict in Fig. \ref{fig_resim} (grey-dotted line)  the simulated RESET obtained  from profile 1  for $V=$-2.65V, which follows the experimental RESET curve almost with the same trend as the IS for $V=$-2.7V.

Thus,  a clever strategy to reduce  the RESET time and, additionally, optimize the power consumption, could be to engineer the initial OV profile \cite{baner_2020}, as   barely different   OV distributions  with quite similar resistance values  give rise  to experimentally accessible lower RESET times for the same applied stimulus or, alternative, similar RESET curves for lower stimulus. \\
The VEOV model was recently extended from 1d to 2d (2d-VEOV), as a tool to  describe how OV electromigration is affected by the presence of  structural or morphological defects, such as dislocations or grain boundaries.
For  the present analysis, we have not taken into account those defects, as we do not have experimental evidence  pointing to any strong influence of such type of morphological defects, neither from  the electrical measurements nor the structural characterization perfomed for this sample \cite{Fer_2020}. However, we analyzed the transient dynamics with the 2d VEOV model
in the $x - y$ plane for different initial OV distributions, when  an external stimulus is applied along the $x$ direction (see Supp. Mat.).
Although the OV evolution during the transient process depends  on the shape and location of the initial OV distribution, after a characteristic time scale an uniform OV profile is attained
along the $y$ direction while  a Gaussian-like profile is obtained in the $x$ direction, for  the studied initial conditions.
The uniformization in the $y$ direction takes place for time scales  much shorter  than the time needed in the experiment for cycling until the stable HSL depicted in Fig. \ref{fig:f2} (a) is obtained. Therefore, under this result,  we  safely   simulate  the  actual RESET process  in terms of many parallel  OV 1d channels along the $x$ direction, employing the 1d version of the  VEOV model, which consumes significantly   less computing time.

\begin{figure}[h!]
    \centering
\includegraphics[width=0.9\linewidth]{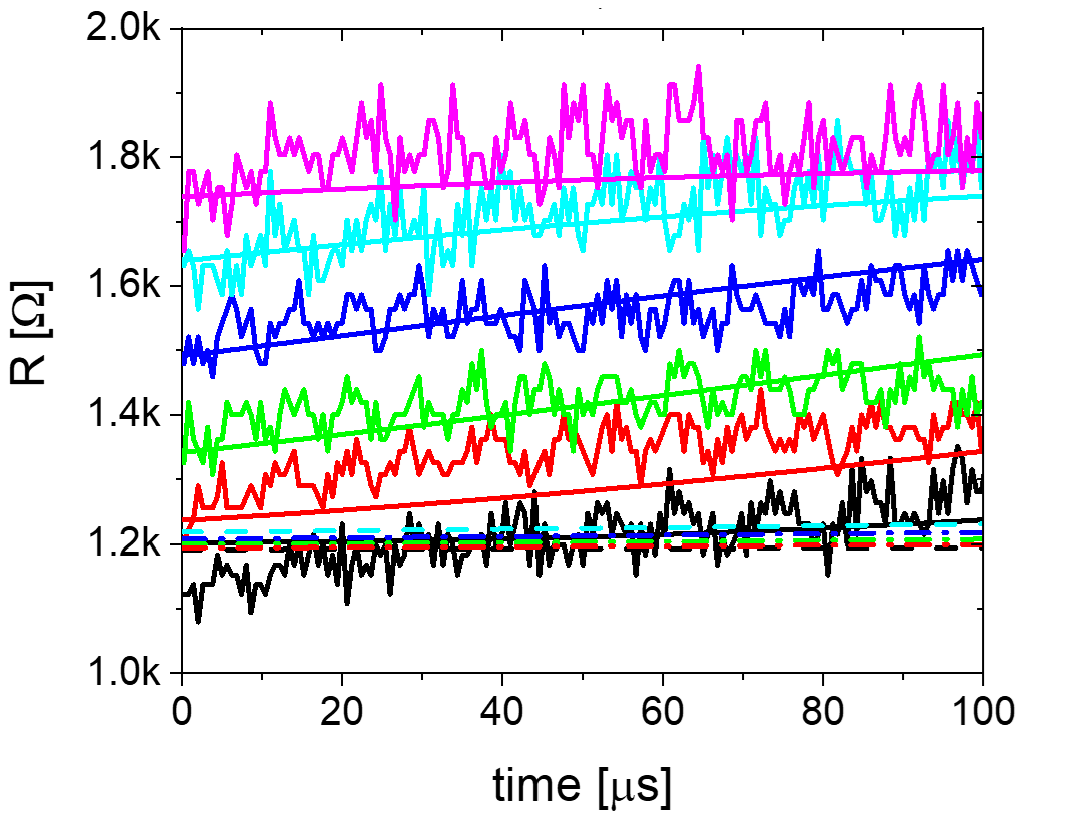}
    \caption{Resistance vs time evolution for six consecutive single pulses of V= -2.4 V and each one  of 100 $\mu$s. The experimental results and the simulated ones- both starting from the IS profile 6 are plotted in solid curves (the simulations correspond to the curves with  no fluctuations). In dashed lines are the simulations for a sequence of  6 pulses of  V= -2.3 V, which altogether do not induce any noticeable change in the resistance.}
    \label{fig:mr}
\end{figure}

As previously discussed, it is clear that reducing the amplitude of the applied voltage could lead to a situation where a  single electrical pulse  could  not be enough to produce a complete RESET, i.e. the device does not  reach its HRS. An interesting  avenue  to explore  is to  apply  many consecutive pulses of lower intensities \cite{ghenzi_2010, ghenzi_2012}, each producing a partial RESET.
In Fig.\ref{fig:mr} we show the experimental evolution   of the remanent resistance for six successive  single pulses of V= -2.4 V with a duration of 100 $\mu$s applied starting from the IS. Each pulse produces a partial RESET that progressively increases  the resistance. In this protocol, the attained resistance  for a given pulse is taken as the initial state for the subsequent pulse. Notice that   once the complete RESET is achieved (in the present case after six pulses),  the HRS resistance  remains constant in time -besides the existence of some experimental fluctuations-.
The simulated partial RESETs reproduce quite  accurately the experimental ones, as shown  in Fig.\ref{fig:mr} with solid lines following the experimental trend. Moreover, when the amplitude of the pulses is reduced to $V=$ -2.3 V, the initial resistance value remains almost unchanged after the sequence of six pulses ended, as it is   shown by the horizontal dashed lines.
We stress that this response persists even after increasing the number of applied pulses. 

Finally, we show now how the already displayed partial RESET transitions can be used to estimate the OV activation energies. If  we  aim to reproduce the multiple partial RESETs using number of pulses $n$ -each one  of 100$\mu s$- given by the experiments, we notice that the VEOV model present two possible control parameters, $-V_0$ and $\Delta V_i$, which are added in the argument of the exponential function that defines the Arrhenius factor that controls the OV transfer rate $p_{i, i\pm1}$.  Lower $V_0$ values (higher $V$ values) imply that fewer pulses are needed to reach the HRS, while higher $V_0$ values (lower $V$ values) require more pulses to achieve the  same HRS. It is noteworthy that the same response could be achieved by adjusting both  parameters, provided that the quantity $(-V_0 + \Delta V_i)$ remains unchanged.
Additionally,  as we have shown, there is an external  threshold voltage $V_{th}$ below which the resistance exhibits no evolution or resistive changes regardless of the number of applied pulses. Above this threshold, the system starts to exhibit resistive changes.  With this threshold voltage we can  define an effective  potential barrier  $V_{{ef}} \equiv (-V_0 + \Delta V_{th})$, where for simplicity we assume a linear voltage drop with  $\Delta V_{th}\equiv V_{th}/N$. 

Thus, given a value of $V_0$, in the simulations we can tune the external voltage in order to determine $V_{{th}}$, and consequently compute $V_{{ef}}$. 
In the present case, we varied the external  voltage  until  for six consecutive pulses the resistive change was below 5$\%$ of full RESET. With this criteria we obtained $V_{{th}}=-2.3V$ for the example depicted by the dashed lines  in Fig. \ref{fig:mr}. 
We assume that  the experimental threshold voltage, $V_{the}$,  is equal to numerical one,  $V_{th}$. Increments of $\pm 0.1$ V are fixed by the accuracy of the pulse generator, imposing an uncertainty in $ V_{the}$.

By increasing the external voltage to $V= -2.4V\equiv V_{6r}$, we   reproduce  the six experimental partial RESETs, depicted by  solid lines in Fig.\ref{fig:mr}, and non-trivial characteristics of the observed behavior. In this context,  we have consider the difference $V_{6r} - V_{{th}}$ as  the experimental uncertainty.

With these ingredients at hand we can  estimate an effective potential barrier for OV electromigration representative of the memristive system under study, by comparing simulations with experimental data.
Following this reasoning, for the present  case we use $V_{the} = $-2.3  V ($\Delta V_{th} \sim 0.02 V$) and from the value of $V_0 = 0.12 V$  employed in the simulations (see Sup. Material) we estimate  e$ V_{efe} = (0.10 \pm 0.01) $eV,  in consistency with reported values \cite{Hur_2019}.

\section {Concluding remarks}

We investigated the RESET process and the associated OV dynamics  for  a
Pt/TaO\(_{2-h}\)/Ta\(_2\)O\(_{5-x}\)/TaO\(_{2-y}\)/Pt  memristive device, using a combination of experimental methods and theoretical simulations based on the VEOV model. We  focus  on the complete RESET process, which involves the transition from LR  to the HR state under electrical stress, providing   insights into the distribution and kinetics of OVs throughout the device during this process. 
An interesting outcome of our analysis is that quite similar initial (low) resistance values  can have associated  different OV profiles the corresponding  RESET transitions are completed in  different times. Thus, it appears that engineering  the OV profiles could be a possible route to change the time scale of this process. 

Additionally, by applying several successive pulses inducing gradual RESETs, we were able to estimate the activation energies associated with OV electromigration, resulting  in very good agreement with reported values \cite{Hur_2019}. This further evidences that the VEOV model is powerful enough to extract relevant  energy scales and system´s parameter values. In addition, we notice that by using the 2d VEOV model it  should be possible to  extend the present study to systems including
different types of extended defects. This will allow determining the influence  of these defects on the memristive behaviour of oxide-based devices. We notice that this could be the key to obtain physical information about the origin of cycle-to-cycle and/or device-to-device variations, which are usually present in RS devices and are often attributed to the present of (uncontrolled) defects, stabilized during the device fabrication process. \cite{baner_2020,bischoff_2022,xu_2023, park_2024}. This topic deserves further consideration and  we leave it for a future work.\\

 We acknowledge support from ANPCyT (PICT2019-02781, PICT2019-0654, PICT2020A-00415), UNCuyo(06/C026-T1) and EU-H2020-RISE project MELON (Grant No. 872631)\\

\textbf{Supplementary Material}

See Supplementary Material for additional information on the electrical characterization, system parameters and for the analysis of the  OV dynamics employing  the 2d VEOV model.

\textbf{AUTHOR DECLARATIONS}

\textbf{Conflict of Interest}

The authors have no conflicts to disclose.



\textbf{Data availability}

The data that support the findings of this study are available from the corresponding author upon reasonable request.

\bibliography{reference}

\end{document}